\documentclass[sigconf]{acmart}
\AtBeginDocument{%
  }

\setcopyright{acmlicensed}
\copyrightyear{2026}
\acmYear{2026}
\setcopyright{cc}
\setcctype{by-nc-nd}
\acmDOI{10.1145/3774905.3794679}
\acmConference[WWW Companion '26]{The Web Conference 2026 Companion}
  {April 13--17, 2026}{Dubai, United Arab Emirates}
\acmBooktitle{Companion Proceedings of the ACM Web Conference 2026 (WWW Companion '26),
  April 13--17, 2026, Dubai, United Arab Emirates}

\acmISBN{979-8-4007-2308-7/2026/04}

\usepackage{algorithm}
\usepackage{algpseudocode}
\algrenewcommand\algorithmicindent{1.1em}
\algrenewcommand\algorithmicrequire{\textbf{Input:}}
\algrenewcommand\algorithmicensure{\textbf{Output:}}
\usepackage{booktabs}
\usepackage{tabularx}
\usepackage{array}
\usepackage{algorithm}
\usepackage{algpseudocode}

\algnewcommand{\algorithmicinput}{\textbf{Input:}}
\algnewcommand{\Input}{\item[\algorithmicinput]}
\algnewcommand{\algorithmicoutput}{\textbf{Output:}}
\algnewcommand{\Output}{\item[\algorithmicoutput]}
\usepackage{placeins}

\usepackage{xcolor}
\usepackage{hyperref}

\hypersetup{
  colorlinks=true,
  linkcolor=blue,   
  citecolor=blue,   
  urlcolor=blue
}

\begin{document}

\title[Privacy-Aware Machine Unlearning for RL-Based Ransomware Detection]{Privacy-Aware Machine Unlearning with SISA for Reinforcement Learning–Based Ransomware Detection}

\author{Jannatul Ferdous}
\email{jferdous@csu.edu.au}
\affiliation{%
  \institution{Charles Sturt University}
  \city{Wagga Wagga}
  \state{NSW}
  \country{Australia}
}

\author{Rafiqul Islam}
\email{mislam@csu.edu.au}
\affiliation{%
  \institution{Charles Sturt University}
  \city{Albury}
  \state{NSW}
  \country{Australia}
}

\author{Md Zahidul Islam}
\email{zislam@csu.edu.au}
\affiliation{%
  \institution{Charles Sturt University}
  \city{Bathurst}
  \state{NSW}
  \country{Australia}
}

\renewcommand{\shortauthors}{Jannatul Ferdous, Rafiqul Islam, and Md Zahidul Islam}

\begin{abstract}
Ransomware detection systems increasingly rely on behavior-based machine learning to address evolving attack strategies. However, emerging privacy compliance, data governance, and responsible AI deployment demand not only accurate detection but also the ability to efficiently remove the influence of specific training samples without retraining the models from scratch. In this study, we present a privacy-aware machine unlearning evaluation framework for reinforcement learning (RL)-based ransomware detection built on Sharded, Isolated, Sliced, and Aggregated (SISA) training. The framework enables efficient data deletion by retraining only the affected model shards rather than the entire detector, substantially reducing the retraining cost while preserving detection performance. We conduct a controlled comparative study using value-based RL agents, including Deep Q-Network (DQN) and Double Deep Q-Network (DDQN), under identical experimental settings with a cost-sensitive reward design and 5-fold stratified cross-validation on Windows~11 behavioral ransomware telemetry. Detection confidence is evaluated using a continuous Q-score margin, enabling ROC--AUC analysis beyond binary predictions. For unlearning, the dataset is partitioned into five shards with majority-vote aggregation, and a fast-unlearning path is evaluated by deleting 5\% of the samples from a single shard and retraining only that shard. Results show that SISA-based unlearning incurs negligible utility degradation ($\leq 0.05\%$ absolute F1 drop) while substantially reducing retraining time relative to full SISA retraining. DDQN exhibits slightly improved stability and lower utility loss than DQN, while both agents maintain near-identical in-distribution performance after unlearning. These findings indicate that SISA provides a practical and computationally efficient unlearning mechanism for RL-based ransomware detection, supporting privacy-aware deployment without compromising security effectiveness.
\end{abstract}




\maketitle

\section{Introduction}
The exponential growth of machine learning (ML) and artificial intelligence (AI) in cybersecurity has fundamentally transformed the way organizations detect and respond to cyber threats~\cite{ref1}. Ransomware has emerged as one of the most destructive and rapidly evolving threats in the digital landscape, with attacks increasing in both frequency and sophistication. Ransomware detection systems increasingly rely on ML to cope with evolving attack strategies; however, they face mounting pressures from privacy regulations, data governance requirements, and responsible AI deployment mandates~\cite{ref2}. While achieving high detection accuracy is critical for security, modern systems must also address an often-overlooked challenge: the ability to efficiently remove the influence of specific training samples without retraining the models from scratch. This need arises from privacy compliance (e.g., GDPR ``right to be forgotten''), data correction requests, and ethical deployment practices~\cite{ref3}.

Machine unlearning has emerged as a technique for selectively removing the influence of designated training samples from trained models without full retraining~\cite{ref4}. Many unlearning strategies remain computationally expensive when deployed in continuously operating security systems. The Sharded, Isolated, Sliced, and Aggregated (SISA) training paradigm is a practical approach that partitions the training data into independent shards, trains separate constituent models, and supports efficient unlearning by retraining only the affected shard from its last valid checkpoint~\cite{ref5}.

While unlearning has been studied extensively for supervised learning and verification settings~\cite{ref6,ref7}, its integration with reinforcement learning (RL), particularly in security-sensitive domains, remains underexplored. To address this gap, this study proposes an integrated framework for privacy-aware machine unlearning designed for RL-based ransomware detection systems. We adopt SISA to enable shard-level retraining and evaluate its interaction with value-based RL agents (DQN and DDQN) under identical experimental settings. Using Windows~11 behavioral ransomware telemetry and a Q-score--based ROC evaluation, we quantify detection performance, utility preservation, and computational overhead before and after one-shard unlearning. The resulting empirical evidence supports privacy-constrained deployment of RL-based ransomware detectors without compromising operational utility.

This study makes four key contributions:
\begin{itemize}
\item We present a systematic study of SISA-based machine unlearning applied to RL-based ransomware detection, enabling efficient unlearning of sensitive training samples without compromising detection performance.

\item We evaluate DQN and DDQN under identical reward designs, hyperparameters, and cross-validation settings to isolate algorithmic effects.

\item We demonstrate efficient one-shard unlearning that achieves negligible utility degradation ($\leq$0.2\% absolute F1) while reducing retraining cost to near-baseline levels.

\item We provide an auditable and deployment-oriented assessment of unlearning behavior aligned with responsible AI requirements in security systems.
\end{itemize}

The remainder of this paper is organized as follows. Section~2 reviews related work on behavior-based ransomware detection, RL in ransomware analytics, and machine unlearning with SISA. Section~3 describes the methodology, including dataset description, RL formulation, Q-score evaluation, and the SISA-based unlearning protocol. Section~4 details the experimental setup, including the Q-network architecture and hyperparameters. Section~5 presents the results. Section~6 discusses implications, limitations, and directions for verifiable and large-scale unlearning. Finally, Section~7 concludes the study.

\section{Related Work}
\label{sec:related}

\subsection{Ransomware Detection via Behavioral Analysis}
Behavior-based ransomware detection has emerged as a robust alternative to static signature methods, which fail in the presence of obfuscation and polymorphism~\cite{ref8}. Early dynamic analysis studies have demonstrated that API calls, file system activity, and registry behavior are effective discriminators for ransomware~\cite{ref9,ref10,ref11}. Subsequent studies have adopted machine learning classifiers and deep neural networks over behavioral traces to improve detection accuracy~\cite{ref12,ref8,ref11,ref13}. Although these approaches achieve high in-distribution performance, they assume immutable training data and do not address post-deployment data removal or privacy constraints.

\subsection{Reinforcement Learning for Cybersecurity and Ransomware}
Recent research has explored reinforcement learning for malware and ransomware detection, motivated by its ability to encode asymmetric security costs and adapt to evolving behavior~\cite{ref14}. Value-based methods, such as DQN and DDQN, have shown promise for structured behavioral telemetry owing to their stability and sample efficiency. DQN has been applied to malware detection and response optimization, whereas Double DQN (DDQN)~\cite{ref15} improves stability by mitigating Q-value overestimation. However, existing RL-based ransomware detectors focus exclusively on detection accuracy~\cite{ref16} or defense mechanisms~\cite{ref17}. The implications of data deletion or unlearning on learned reinforcement learning (RL) policies remain largely unexplored.

\subsection{Machine Unlearning and SISA}
Machine unlearning addresses regulatory requirements such as the GDPR right to erasure by enabling models to forget specific training samples~\cite{ref3,ref18}. Among scalable approaches, SISA (Sharded, Isolated, Sliced, and Aggregated) training allows efficient unlearning by retraining only the affected data shards~\cite{ref5,ref19}. Prior studies have validated SISA primarily in supervised learning settings and evaluated its computational efficiency and privacy guarantees~\cite{ref6,ref7}. Its application to reinforcement learning--based security systems, particularly ransomware detection, has received little attention.

\subsection{Gap Addressed by This Study}
To the best of our knowledge, no prior study has systematically evaluated SISA-based unlearning in reinforcement-learning-based ransomware detection. Existing studies have focused on RL detection without unlearning or on unlearning mechanisms without considering RL-specific evaluation, cost-sensitive rewards, or security-critical constraints. This study bridges these gaps by evaluating DQN and DDQN under identical settings and quantifying the impact of one-shard SISA unlearning on detection utility and runtime. This evaluation addresses the gap between adaptive ransomware detection and responsible artificial intelligence (AI) deployment.

\section{Methodology}
\label{sec:methodology}

This section presents a privacy-aware reinforcement learning framework for ransomware detection using SISA-based machine unlearning. The proposed methodology supports efficient data deletion through sharded retraining, in line with responsible AI requirements.

\subsection{Dataset and Feature Representation}
\label{sec:dataset}

Experiments were conducted on our custom behavioral ransomware dataset comprising 2000 executables (1000 ransomware and 1000 benign). Ransomware samples span 30 high-impact families selected based on multi-year (2019--2024) threat-intelligence prevalence and multi-vendor verification, sourced from MalwareBazaar and VirusShare, while benign samples were collected from trusted repositories (SnapFiles, PortableApps, and GitHub). Individual ransomware samples were retained only if at least 45 VirusTotal engines flagged them as malicious, at least 15 explicitly labeled them as ransomware, and at least 10 engines agreed on family attribution.

Samples were executed in the ANY.RUN Windows~11 sandbox to ensure high-fidelity behavioral capture and avoid legacy Cuckoo limitations, producing JSON reports with approximately 11k raw indices and around 250 behavioral fields. To ensure reproducibility and prevent data leakage, preprocessing steps, including biased entry removal, categorical encoding, feature scaling, and standardization, were performed to create an optimized dataset with selected ransomware features covering filesystem, registry, process, API, network, CryptoAPI, incident rules, reputation scores, and multi-process behaviors. The final output was a 103-dimensional feature vector suitable for ML- and RL-based ransomware detection.

\subsection{Problem Formulation}
\label{sec:problem}

Let the labeled behavioral dataset be
\begin{equation}
\mathcal{D} = \left\{ (\mathcal{X}_i, \mathcal{Y}_i) \right\}_{i=1}^{N}, 
\quad \mathcal{Y}_i \in \{0,1\},
\label{eq:dataset}
\end{equation}
where $N=2000$, $\mathcal{Y}_i = 1$ denotes ransomware and $\mathcal{Y}_i = 0$ denotes benign behavior.

The objective is to learn the detection function
\begin{equation}
f(\mathcal{X}) \rightarrow \{0,1\},
\label{eq:detector}
\end{equation}
that maximizes detection accuracy while enabling post-training deletion of selected samples without full retraining.

\subsection{Reinforcement Learning Environment Design}
\label{sec:rl_env}

Ransomware detection is modelled as a binary decision-making problem using reinforcement learning, where each training instance corresponds to a state, an action, and a reward transition.

\paragraph{State.}
Each behavioural feature vector corresponds to a state,
\begin{equation}
s \in \mathbb{R}^{d},
\label{eq:state}
\end{equation}
where $d$ is the number of features ($d=103$).

\paragraph{Action.}
The agent (Q-network) learns to classify behavioural states using two discrete actions:
\begin{equation}
a \in \{0,1\},
\label{eq:action}
\end{equation}
where $0$ denotes benign and $1$ denotes ransomware, respectively.

\paragraph{Reward function (cost-sensitive).}
To reflect the asymmetric risk in ransomware detection, all experiments used a fixed cost-sensitive reward function that penalizes false negatives more severely than false positives. The reward is defined as
\begin{equation}
r(y,a)=
\begin{cases}
+1, & a=y,\\
-2, & y=1 \wedge a=0 \quad \text{(false negative)},\\
-0.5, & y=0 \wedge a=1 \quad \text{(false positive)},
\end{cases}
\label{eq:reward}
\end{equation}
where $y$ denotes the ground-truth label. This configuration reflects standard security practice, where undetected ransomware poses substantially greater risk than benign misclassification.

\subsection{Value-Based Deep Reinforcement Learning Algorithms}
\label{sec:value_based}

This study focuses on value-based reinforcement learning methods because of their stable optimization behaviour and explicit state--action value estimation when applied to structured behavioural telemetry. In addition, value-based methods provide action--value margins, which are exploited for confidence ranking and Q-score--based ROC analysis under privacy-driven deletion constraints.

Two value-based deep reinforcement learning algorithms were evaluated:
\begin{itemize}
  \item Deep Q-Network (DQN)
  \item Double Deep Q-Network (DDQN)
\end{itemize}

The DQN and DDQN shared identical experimental settings, including the network architecture, replay buffer size, learning rate, exploration schedule, discount factor, training steps, and random seeds. The only difference lies in the temporal-difference (TD) target computation. This ensures that performance differences can be attributed to the value estimation mechanism rather than confounding hyperparameters.

Both DQN and DDQN learn an action--value function $Q_{\theta}(s,a)$ parameterized by a neural network with weights $\theta$. Learning proceeds by minimizing the TD loss over transitions sampled from an experience replay buffer. Specifically, the optimization objective is
\begin{equation}
\mathcal{L}(\theta)=
\mathbb{E}_{(s,a,r,s') \sim \mathcal{D}}
\left[
\left( y - Q_{\theta}(s,a) \right)^2
\right],
\label{eq:td_loss}
\end{equation}
where $\mathcal{D}$ denotes the replay buffer and $y$ is a TD target computed using a separate target network. The discount factor $\gamma \in (0,1)$ controls the weighting of future rewards.

In DQN, the TD target is computed by directly maximizing the target network's estimated action value at the next state:
\begin{equation}
y^{\mathrm{DQN}} = r + \gamma \max_{a'} Q_{\theta^-}(s',a').
\label{eq:dqn_target}
\end{equation}
Although effective, this formulation is known to suffer from overestimation bias, as the same value function is used for both action selection and evaluation.

DDQN mitigates the overestimation bias by decoupling action selection from action evaluation. The online network selects the next action, whereas the target network evaluates its value:
\begin{equation}
y^{\mathrm{DDQN}} =
r + \gamma \,
Q_{\theta^-}\!\left(
s',\,
\arg\max_{a'} Q_{\theta}(s',a')
\right).
\label{eq:ddqn_target}
\end{equation}
Here, $Q_{\theta^-}$ denotes the frozen target network used for value evaluation, and $Q_{\theta}$ denotes the online network used for action selection. This modification improves stability without increasing computational complexity, making DDQN particularly suitable for structured behavioural telemetry.

\subsection{Cross-Validation and Training Protocol}
\label{sec:cv_protocol}

All experiments were conducted using 5-fold stratified cross-validation with a fixed random seed to ensure that performance differences arose solely from algorithmic behaviour rather than configuration bias. For each fold, the RL agent is trained on 80\% of the data, evaluated on the held-out 20\%, and results are aggregated across folds. Out-of-fold (OOF) predictions were used to compute confusion matrices, ROC curves, and summary statistics.

\subsection{Q-Score--Based ROC Evaluation}
\label{sec:qscore_roc}

Reinforcement learning agents do not generate calibrated probabilities. To enable ROC analysis, we derived a continuous confidence score from the learned action--value function, called the \emph{Q-score}. For each sample $\mathcal{X}$, the Q-score is defined as
\begin{equation}
\text{Q-score}(\mathcal{X}) =
Q_{\theta}(\mathcal{X}, a=1) - Q_{\theta}(\mathcal{X}, a=0).
\label{eq:qscore}
\end{equation}
The margin indicates confidence in predicting ransomware versus benign behaviour. ROC curves and AUC values were computed using Q-scores rather than discrete actions, providing a faithful assessment of ranking performance and confidence.

\subsection{SISA-Based Unlearning with Ensemble Aggregation}
\label{sec:sisa_unlearning}

Machine unlearning aims to remove the influence of specific training samples from deployed models in response to privacy regulations, data correction requirements, or security policies. In learning-based security systems, full retraining after every deletion request is often computationally infeasible, particularly under continuous operational constraints.

\begin{figure}[h]
  \centering
  \includegraphics[width=\linewidth]{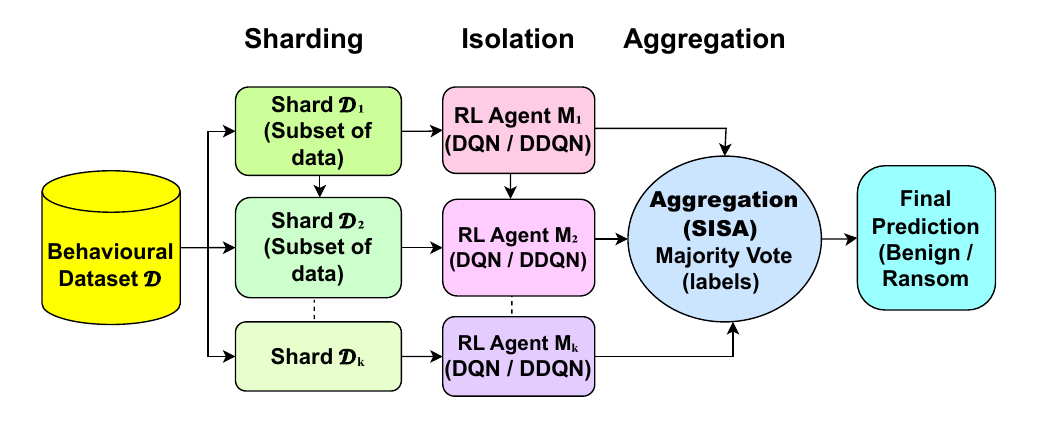}
  \caption{Workflow of the SISA-based training and aggregation framework.}
  \label{fig:sisa_workflow}
\end{figure}

To enable efficient unlearning, we adopt the Sharded, Isolated, Sliced, and Aggregated (SISA) training paradigm. As illustrated in Figure~\ref{fig:sisa_workflow}, SISA partitions the training dataset into multiple disjoint shards, trains independent models on each shard in isolation, and aggregates predictions at inference time. This design confines the influence of individual samples to a limited subset of models, enabling targeted retraining.

In dataset partitioning (sharding phase), the training dataset is partitioned into $M=5$ disjoint shards, and an independent reinforcement learning agent (DQN or DDQN) is trained on each shard. Each shard-level agent learns an action--value function using only its local subset of data, thereby ensuring isolation between shards.

At inference time, each agent produces a class prediction based on its learned Q-function:
\begin{equation}
\hat{\mathcal{Y}}_{m}(\mathcal{X}) = \arg\max_{a} Q_{m}(\mathcal{X}, a),
\label{eq:shard_pred}
\end{equation}
where $Q_{m}$ denotes the action--value function learned by the $m$-th shard agent. These shard-level predictions correspond to the isolation stage shown in Fig.~\ref{fig:sisa_workflow}, where each RL agent independently evaluates the input sample.

The final classification is obtained via majority voting across all shard-level predictions:
\begin{equation}
\hat{\mathcal{Y}}_{\mathrm{SISA}}(\mathcal{X}) =
\mathbb{I}\!\left[
\frac{1}{M}\sum_{m=1}^{M} \hat{\mathcal{Y}}_{m}(\mathcal{X}) \ge 0.5
\right],
\label{eq:majority_vote}
\end{equation}
where $\mathbb{I}[\cdot]$ denotes the indicator function. This ensemble aggregation improves robustness while preserving shard isolation.

When a deletion request is issued, only the shard(s) containing the affected samples are retrained, whereas all other shard models remain unchanged. In this study, we evaluate a fast-unlearning path in which 5\% \ of the samples of a single shard are removed and only that shard is retrained. This approach substantially reduces unlearning cost while preserving predictive utility. Overall, the SISA framework balances computational efficiency and deletion fidelity, making it suitable for privacy-aware ransomware detection systems.

\subsection{Utility Preservation Measurement}
\label{sec:utility_preservation}
To quantify the impact of privacy-driven unlearning on model performance, we measured utility preservation using the change in F1-score before and after unlearning. The utility drop is defined as
\begin{equation}
\Delta \mathrm{F1} = \left| \mathrm{F1}_{\text{before}} - \mathrm{F1}_{\text{after}} \right|.
\label{eq:delta_f1}
\end{equation}
This absolute measure provides a non-negative and symmetric estimate of performance change, where smaller $\Delta \mathrm{F1}$ indicates stronger utility preservation.

Algorithm 1 summarizes the evaluation pipeline, covering value-based RL training, Q-score–based detection, SISA unlearning, and utility preservation under identical experimental settings.

\FloatBarrier
\begin{algorithm*}[t]
\caption{Privacy-Aware Evaluation of Value-Based Reinforcement Learning with SISA Fast Unlearning}
\label{alg:sisa-rl}

\begingroup
\linespread{0.96}\selectfont            

\begin{algorithmic}[1]
\setlength{\itemsep}{0.45ex}            
\setlength{\topsep}{0.6ex}              
\setlength{\partopsep}{0pt}
\setlength{\parsep}{0pt}

\Input Behavioral dataset $D=\{(x_i,y_i)\}_{i=1}^{N}$ with $d{=}103$ features; algorithms $\{\mathrm{DQN},\mathrm{DDQN}\}$; folds $K{=}5$; steps $T$; discount $\gamma$; reward $R2$; shards $M$; forget fraction $f$.
\Output Cross-validated ID performance, Q-score ROC--AUC, runtime cost, and utility preservation before/after unlearning.

\State Split $D$ into $K$ stratified folds
\For{each algorithm $A \in \{\mathrm{DQN},\mathrm{DDQN}\}$}
  \State Initialize $\mathrm{OOF}\_\mathrm{true} \leftarrow \emptyset$, $\mathrm{OOF}\_\mathrm{pred} \leftarrow \emptyset$, $\mathrm{OOF}\_\mathrm{qscore} \leftarrow \emptyset$
  \For{$k \gets 1$ to $K$}
    \State $(D_{\mathrm{tr}},D_{\mathrm{te}}) \leftarrow \mathrm{StratifiedSplit}(D,k)$
    \State $(X_{\mathrm{tr}},X_{\mathrm{te}}) \leftarrow \mathrm{Standardize}(D_{\mathrm{tr}}.X,\ D_{\mathrm{te}}.X)$

    \Statex \vspace{0.6ex}\textit{Base RL training}
    \State Initialize $Q_{\mathrm{online}}$, $Q_{\mathrm{target}}$, and replay buffer $\mathcal{B}$
    \For{$t \gets 1$ to $T$}
      \State $s \leftarrow X_{\mathrm{tr}}[t \bmod |X_{\mathrm{tr}}|]$
      \State $s' \leftarrow X_{\mathrm{tr}}[(t{+}1)\bmod |X_{\mathrm{tr}}|]$
      \State $a \leftarrow \epsilon\text{-}\mathrm{greedy}(Q_{\mathrm{online}}, s)$
      \State $r \leftarrow R2(y(s), a)$
      \State Store transition $(s,a,r,s')$ in $\mathcal{B}$
      \If{$|\mathcal{B}| \ge \mathrm{batch\_size}$}
        \If{$A=\mathrm{DDQN}$}
          \State $a^\star \leftarrow \arg\max_{\tilde a} Q_{\mathrm{online}}(s',\tilde a)$
          \State $y_{\mathrm{TD}} \leftarrow r + \gamma \cdot Q_{\mathrm{target}}(s',a^\star)$
        \Else
          \State $y_{\mathrm{TD}} \leftarrow r + \gamma \cdot \max_{\tilde a} Q_{\mathrm{target}}(s',\tilde a)$
        \EndIf
        \State Update $Q_{\mathrm{online}}$ using $y_{\mathrm{TD}}$
      \EndIf
      \State Periodically update $Q_{\mathrm{target}}$
    \EndFor

    \Statex \vspace{0.6ex}\textit{In-distribution evaluation}
    \State $\hat{y} \leftarrow \arg\max_a Q_{\mathrm{online}}(X_{\mathrm{te}}, a)$
    \State $q \leftarrow Q_{\mathrm{online}}(X_{\mathrm{te}},1) - Q_{\mathrm{online}}(X_{\mathrm{te}},0)$
    \State Append $y_{\mathrm{te}}$, $\hat{y}$, and $q$ to OOF sets

    \Statex \vspace{0.6ex}\textit{SISA training and fast unlearning}
    \State Partition $D_{\mathrm{tr}}$ into $M$ disjoint shards
    \State Train one RL agent per shard (isolated shard models)
    \State $\hat{y}^{\mathrm{SISA}}_{\mathrm{before}} \leftarrow \mathrm{MajorityVote}(\text{shard\_models}, X_{\mathrm{te}})$
    \State $F1_{\mathrm{before}} \leftarrow F1(y_{\mathrm{te}}, \hat{y}^{\mathrm{SISA}}_{\mathrm{before}})$
    \State Select shard $m^\star$ and deletion set $F \subset m^\star$ with $|F|=\lceil f\cdot|m^\star|\rceil$
    \State Retrain only shard $m^\star$ on retained data
    \State $\hat{y}^{\mathrm{SISA}}_{\mathrm{after}} \leftarrow \mathrm{MajorityVote}(\text{updated\_models}, X_{\mathrm{te}})$
    \State $F1_{\mathrm{after}} \leftarrow F1(y_{\mathrm{te}}, \hat{y}^{\mathrm{SISA}}_{\mathrm{after}})$
    \State $\Delta F1 \leftarrow |F1_{\mathrm{before}} - F1_{\mathrm{after}}|$
    \State Record runtime and utility metrics
  \EndFor

  \State Aggregate fold-wise results
  \State Compute OOF confusion matrix and Q-score ROC--AUC
\EndFor
\end{algorithmic}
\endgroup
\end{algorithm*}

\section{Experimental Setup}
All experiments were conducted in a CPU-only Google Colab Pro environment with 51 GB RAM (Cloud usable), ensuring that the proposed framework remains practical for real-world deployment without GPU dependence. The implementation was developed in Python~3.10 using PyTorch~2.1 and scikit-learn~1.3. This setup reflects realistic enterprise constraints and avoids reliance on specialized hardware accelerators. The architectural design and training hyperparameters are reported in Tables~\ref{tab:qnet_arch} and~\ref{tab:train_hparams}, respectively.

\begin{table}[h]
\caption{Q-Network Architecture.}
\label{tab:qnet_arch}
\centering
\small
\begin{tabular}{@{}p{0.30\columnwidth}p{0.64\columnwidth}@{}}
\toprule
\textbf{Layer} & \textbf{Configuration} \\
\midrule
Input & 103 neurons (behavioral feature vector) \\
Hidden Layer 1 & 128 neurons, ReLU activation \\
Hidden Layer 2 & 128 neurons, ReLU activation \\
Output & 2 neurons (Q-values for class 0: benign, class 1: ransomware) \\
\bottomrule
\end{tabular}
\end{table}

\begin{table}[h]
\caption{Training Hyperparameters.}
\label{tab:train_hparams}
\centering
\small
\begin{tabular}{@{}p{0.45\columnwidth}p{0.49\columnwidth}@{}}
\toprule
\textbf{Hyperparameter} & \textbf{Value} \\
\midrule
Optimizer & Adam, learning rate $\alpha=0.001$ \\
Batch size & 64 \\
Training budget & 10{,}000 timesteps per agent (streaming updates) \\
Discount factor & $\gamma=0.1$ \\
Replay buffer size & 50{,}000 \\
Target network update & Every 500 training steps \\
Exploration strategy & $\epsilon$-greedy, $\epsilon: 1.0 \rightarrow 0.05$ (linear decay over 5{,}000 steps) \\
Loss function & Smooth L1 (Huber) loss \\
Reward function & Cost-sensitive (R2): +1 (correct), $-2$ (false negative), $-0.5$ (false positive) \\
\bottomrule
\end{tabular}
\end{table}

\section{Results}
This section reports the experimental results of the proposed privacy-aware reinforcement learning framework with SISA-based machine unlearning. The evaluation focuses on (i) baseline detection performance, (ii) utility preservation before and after one-shard SISA unlearning, and (iii) computational efficiency and retraining overhead.

\subsection{Baseline Detection Performance}
This subsection presents the in-distribution (ID) ransomware detection performance of the baseline value-based reinforcement learning agents before any unlearning is applied. Performance is reported using the F1-score, worst-fold F1, and Q-score--based ROC--AUC, alongside the average training and inference time per fold.

Table~\ref{tab:id_baseline} summarizes the 5-fold cross-validated in-distribution ransomware detection performance of baseline value-based reinforcement learning agents. Both DQN and DDQN achieved near-perfect detection accuracy, with mean F1-scores above 0.99 and Q-score ROC--AUC values exceeding 0.998, demonstrating a strong discriminative capability for behavioral telemetry. The DDQN marginally outperformed the DQN in terms of the mean F1-score and exhibits improved worst-fold stability, consistent with its reduced overestimation bias in value estimation.

\begin{table*}[t]
\centering
\caption{In-Distribution Detection Performance (Baseline RL Agents).}
\label{tab:id_baseline}
\renewcommand{\arraystretch}{1.15}
\begin{tabular}{lcccccc}
\hline
Model &
ID F1 ($\mu \pm \sigma$) &
Worst-fold F1 &
OOF AUC (Q-score) &
Training time (s) &
Inference time (s) \\
\hline
DQN  & 0.9920 $\pm$ 0.0045 & 0.9850 & 0.9987 & 23.14 & 0.0365 \\
DDQN & 0.9925 $\pm$ 0.0025 & 0.9900 & 0.9983 & 25.13 & 0.0457 \\
\hline
\end{tabular}
\end{table*}

\begin{figure}[h]
  \centering
  \includegraphics[width=\columnwidth]{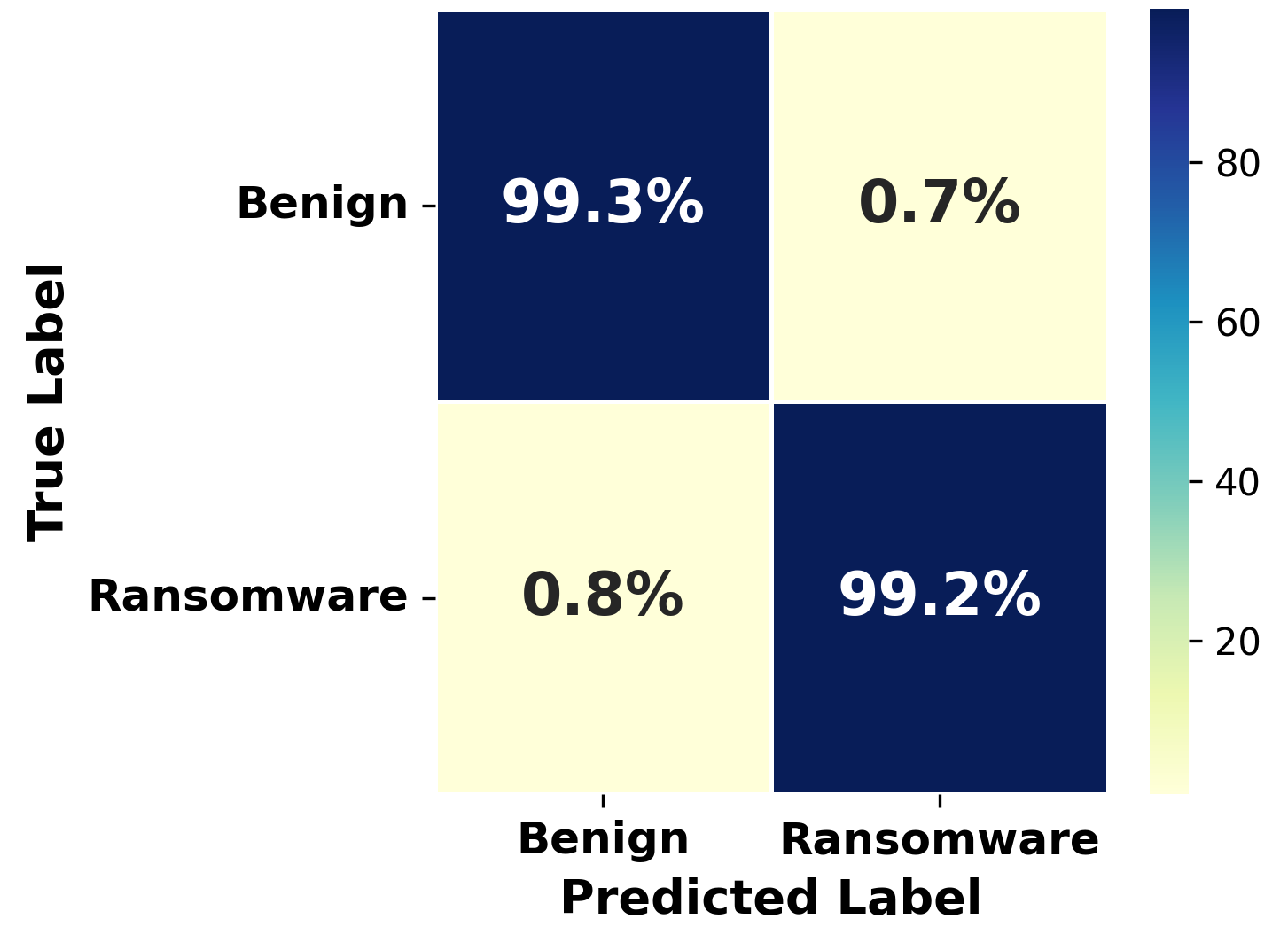} 
  \caption{Confusion Matrix for DDQN Baseline Detection.}
  \label{fig:cm_ddqn}
\end{figure}

Figure~\ref{fig:cm_ddqn} further analyzes the winning model (DDQN) using an out-of-fold confusion matrix. The model correctly classified 99.3\% of benign samples and 99.2\% of ransomware samples, with false-positive and false-negative rates limited to 0.7\% and 0.8\%, respectively. This balanced error profile confirms that the cost-sensitive reward formulation effectively prioritizes ransomware detection while maintaining low benign misclassification rates.

\begin{figure}[h]
  \centering
  \includegraphics[width=\columnwidth]{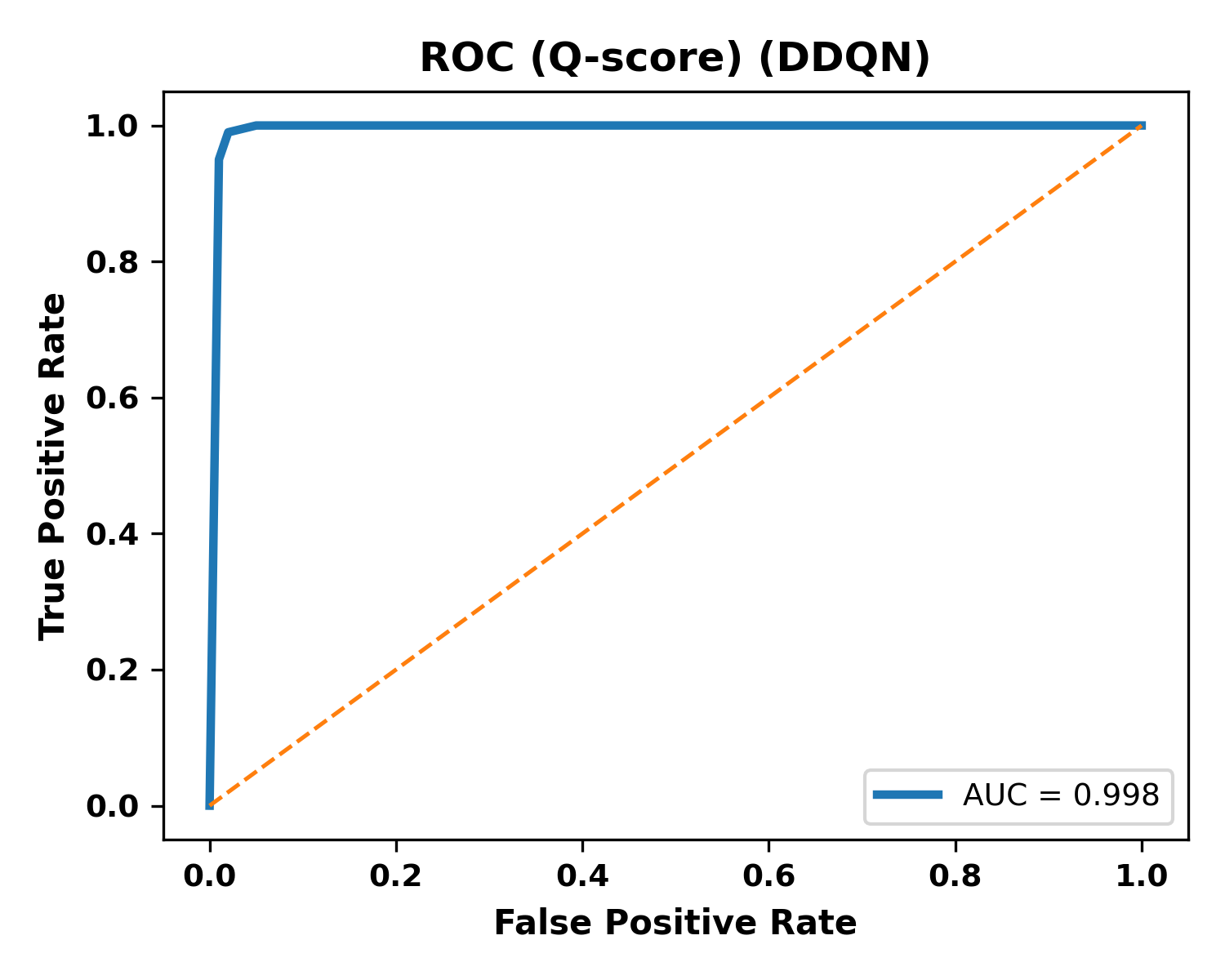} 
  \caption{Q-Score ROC Curve for DDQN Baseline Detection.}
  \label{fig:roc_ddqn}
\end{figure}

Figure~\ref{fig:roc_ddqn} shows the ROC curve derived from the proposed Q-score margins. The DDQN achieved an OOF ROC--AUC of approximately 0.998, indicating a near-perfect ranking performance. By leveraging action-value differences rather than calibrated probabilities, this evaluation provides a faithful assessment of confidence and decision robustness for reinforcement learning--based detectors. Overall, these results establish the DDQN as a stable and reliable baseline, forming a strong reference point for subsequent privacy-driven SISA unlearning analyses.

\subsection{Privacy-Driven Utility Preservation under SISA Unlearning}
To evaluate the impact of privacy-driven unlearning, we apply SISA-based sharded retraining with $M=5$ shards, followed by a fast-unlearning path, in which 5\% of samples are removed from a single shard, and only the affected shard is retrained. Table~\ref{tab:utility_sisa} reports the detection performance before and after unlearning, which was measured using ensemble majority-vote predictions on the test set.

\begin{table}[h]
\caption{Utility Preservation Before and After One-Shard SISA Unlearning.}
\label{tab:utility_sisa}
\centering
\small
\begin{tabular}{@{}lccc@{}}
\toprule
\textbf{Model} & \textbf{SISA F1 (Before)} & \textbf{SISA F1 (After)} & \textbf{Utility Drop ($\Delta$F1)} \\
\midrule
DQN  & 0.9787 & 0.9782 & 0.0005 \\
DDQN & 0.9806 & 0.9806 & 0.0000 \\
\bottomrule
\end{tabular}
\end{table}

Across both DQN and DDQN, the observed utility degradation was negligible. The average F1-score drop was approximately 0.05\% for DQN and 0.00\% for DDQN, indicating that one-shard unlearning preserves almost all detection capabilities. These results confirm that SISA-based unlearning enables practical data deletion with a minimal impact on ransomware detection effectiveness, satisfying privacy requirements without compromising operational security.

\subsection{Computational Efficiency and Retraining Overhead}
Table~\ref{tab:cost_comp} summarizes the average computational costs across five folds for both the DQN and DDQN. Full SISA training increases the cost by approximately fivefold relative to the baseline training, reflecting the overhead of maintaining shard isolation. However, the one-shard unlearning strategy retrains only the affected shard, reducing the unlearning time to near-baseline levels ($\approx$22--24~s). This confirms that SISA-based unlearning enables efficient, scalable, and privacy-compliant model updates without incurring the prohibitive costs of full retraining.

\begin{table}[h]
\centering
\caption{Computational Cost Comparison: Baseline Training vs.\ SISA Unlearning.}
\label{tab:cost_comp}
\resizebox{\columnwidth}{!}{%
\begin{tabular}{lccc}
\hline
Model &
Baseline Train (s) &
SISA Full Train (s) &
One-Shard Unlearn (s) \\
\hline
DQN  & 23.14 & 113.30 & 22.40 \\
DDQN & 25.13 & 123.21 & 23.98 \\
\hline
\end{tabular}}
\end{table}

\begin{figure}[h]
  \centering
  \includegraphics[width=\columnwidth]{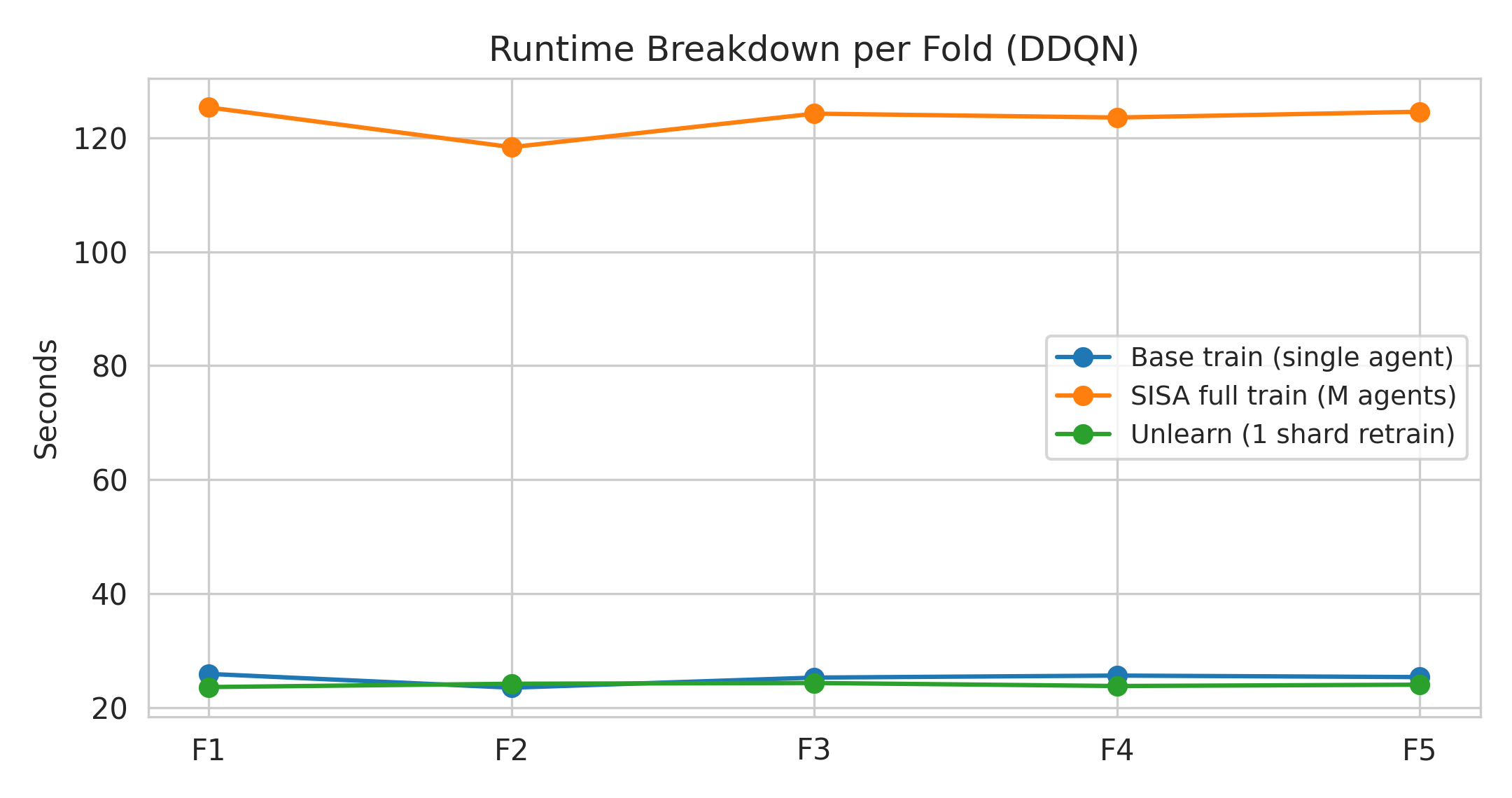} 
  \caption{Runtime Breakdown per Fold (DDQN).}
  \label{fig:runtime_ddqn}
\end{figure}

Figure~\ref{fig:runtime_ddqn} illustrates the fold-wise runtime breakdown for the DDQN model, comparing the baseline single-agent training, full SISA training with five shard-level agents, and one-shard retraining for privacy-driven unlearning purposes. Baseline training remains stable across folds ($\approx$25~s), whereas full SISA training incurs a higher cost owing to independent shard-wise model replication. In contrast, one-shard retraining consistently requires a runtime comparable to that of baseline training, demonstrating the effectiveness of the proposed fast-unlearning path.

This reduction in retraining overhead enables continuous operation while supporting privacy-driven deletion, essential for real-world ransomware detection systems subject to regulatory and operational constraints.

\section{Discussion}
This study examines whether privacy-aware machine unlearning can be incorporated into reinforcement learning--based ransomware detection without significantly affecting detection performance. The results show that the SISA-enabled DDQN model preserves near-baseline in-distribution accuracy while enabling efficient shard-level deletion with substantially lower retraining costs than full-model retraining. This suggests that unlearning can be integrated into security-focused RL systems with limited operational disruptions.

The findings are consistent with earlier work demonstrating the effectiveness of value-based reinforcement learning for behavioral malware detection under cost-sensitive reward structures. However, most existing approaches assume static training data and do not address post-training deletion requirements. By introducing shard-isolated retraining, this study extends prior methods to scenarios in which data removal is required after deployment. The observed stability of DDQN further indicates that reduced value overestimation may help preserve model behavior following partial retraining.

This study focuses on value-based reinforcement learning rather than policy-gradient or actor--critic methods because the evaluation relies on explicit state--action value estimates. In particular, the Q-score margin $Q(\mathcal{X},1) - Q(\mathcal{X},0)$ is used for confidence ranking and ROC analysis, which cannot be directly obtained from stochastic policy outputs. Therefore, value-based methods provide interpretable decision margins under privacy-driven deletion constraints, making them well suited to the security-oriented evaluation setting considered in this study.

Although the evaluation is limited to single-shard deletions and does not include formal verification of forgetting, the results indicate that practical, privacy-aware unlearning can be integrated into security systems with minimal operational impact. This study represents an initial step toward responsible deployment of RL-based ransomware detection and motivates future studies on verifiable and large-scale unlearning.

\section{Conclusion and Future Work}
This study presents a privacy-aware reinforcement learning framework for ransomware detection that integrates SISA-based machine unlearning into value-based RL agents. Through a controlled comparative evaluation of DQN and DDQN under identical experimental settings on Windows~11 behavioral ransomware telemetry, we demonstrated that one-shard SISA unlearning enables efficient data deletion with negligible impact on detection performance, while substantially reducing retraining overhead compared to full-shard retraining. The results confirm that DDQN provides stable and robust behavioral detection, achieving near-perfect in-distribution performance ($\mathrm{F1} > 0.99$, $\mathrm{AUC} > 0.998$), and that privacy-driven unlearning can be achieved without compromising baseline security guarantees. Overall, these findings establish SISA as a practical, auditable, and computationally efficient unlearning mechanism for RL-based ransomware detection, supporting responsible AI deployment in security-critical applications.

Future research will extend SISA-based unlearning to multi-shard and sequential deletion scenarios, targeted sensitive-sample removal, and evaluation under broader datasets and adversarial conditions. In addition, we will investigate policy-level and actor--critic methods under SISA-based unlearning as a further research direction. Finally, we will examine verifiable unlearning mechanisms, including oracle-based forgetting tests and membership inference, to provide stronger evidence that the model has genuinely forgotten deleted data and to enhance the credibility of privacy guarantees.


\bibliographystyle{ACM-Reference-Format}
\bibliography{references}

\end{document}